\documentclass[	
						prl,%
						superscriptaddress,%
						twocolumn,%
						a4paper,
						showkeys,
					]{revtex4}

\usepackage[frenchb,english]{babel}
\usepackage[T1]{fontenc}
\usepackage{upgreek}
\usepackage{textcomp} 				
\usepackage[dvips]{graphicx}		
\usepackage{color}					
\usepackage{amssymb,amsmath}		
\usepackage{array}					
\usepackage{ulem}

\begin{document}

\title{Jahn-Teller, Polarity and Insulator-to-Metal Transition in BiMnO$_3$ at High Pressure}
\date{\today}
\author{Mael Guennou}
\email{guennou@lippmann.lu}
\affiliation{Département Science et Analyse des Matériaux, CRP Gabriel Lippmann, 41 rue du Brill, Belvaux, L-4422 Luxembourg}
\author{Pierre Bouvier}
\email{pierre.bouvier@grenoble-inp.fr}
\affiliation{Laboratoire des Mat\'eriaux et du G\'enie Physique, CNRS, Universit\'e Grenoble-Alpes, 3 Parvis Louis N\'eel, 38016 Grenoble, France}
\affiliation{European Synchrotron Radiation Facility (ESRF), BP 220, 6 Rue Jules Horowitz, 38043 Grenoble Cedex, France}
\author{Pierre Toulemonde}
\email{pierre.toulemonde@neel.fr}
\author{C\'eline Darie}
\author{C\'eline Goujon}
\author{Pierre Bordet}
\affiliation{Universit\'e Grenoble-Alpes, Institut N\'eel, F-38042 Grenoble, France\\CNRS, Institut N\'eel, F-38042 Grenoble, France}
\author{Michael Hanfland}
\affiliation{European Synchrotron Radiation Facility (ESRF), BP 220, 6 Rue Jules Horowitz, 38043 Grenoble Cedex, France}
\author{Jens Kreisel}
\affiliation{Département Science et Analyse des Matériaux, CRP Gabriel Lippmann, 41 rue du Brill, Belvaux, L-4422 Luxembourg}

\begin{abstract}
The interaction of coexisting structural instabilities in multiferroic materials gives rise to intriguing coupling phenomena and extraordinarily rich phase diagrams, both in bulk materials and strained thin films. Here we investigate the multiferroic BiMnO$_3$ with its peculiar 6$s^2$ electrons and four interacting mechanisms: electric polarity, octahedra tilts, magnetism, and cooperative Jahn-Teller distortion. We have probed structural transitions under high pressure by synchrotron x-ray diffraction and Raman spectroscopy up to 60~GPa. We show that BiMnO$_3$ displays under pressure a rich sequence of five phases with a great variety of structures and properties, including a metallic phase above 53~GPa and, between 37 and 53~GPa, a strongly elongated monoclinic phase that allows ferroelectricity, which contradicts the traditional expectation that ferroelectricity vanishes under pressure. Between 7 and 37~GPa, the $Pnma$ structure remains remarkably stable but shows a reduction of the Jahn-Teller distortion in a way that differs from the behavior observed in the archetypal orthorhombic Jahn-Teller distorted perovskite LaMnO$_3$. 
\end{abstract}

\keywords{BiMnO$_3$, Multiferroics, X-ray diffraction, High pressure, Raman spectroscopy, Perovskite}

\maketitle

Functional properties of $AB$O$_3$ perovskites are intimately linked to their structural distortions away from the ideal cubic perovskite, which can be schematically decomposed into rotations (tilts) of $B$O$_6$ octahedra, (anti)polar cation shifts and distortions of octahedra, typically induced by a cooperative Jahn-Teller effect and orbital ordering \cite{Mitchell2002,Lufaso2004,Carpenter2009}. A perovskite at given external conditions may exhibit one or a combination of these distortions. Their respective stability, their evolution and coupling as a function of temperature, pressure or chemical substitution give rise to a large variety of crystal structures and an equally rich variety of physical properties \cite{Mitchell2002}. The family of bismuth-based perovskites displays a particularly remarkable diversity of structures \cite{Belik2012}, and the peculiar chemistry related to the 6$s^{2}$ lone pair electrons of the Bi$^{3+}$ and the associated polar properties are of great interest in the search for new lead-free piezoelectrics \cite{Roedel2009,Keeble2013}. The by far most studied is the multiferroic bismuth ferrite BiFeO$_3$ (BFO), which presents at 300~K a combination of strong octahedra tilts, strong ferroelectricity and canted antiferromagnetism \cite{Catalan2009}. The interplay between these instabilities gives rise to a rich and not fully understood $P$--$T$ phase diagram \cite{Catalan2009,Guennou2011a} with six phase transitions between 0 and 60~GPa at room temperature, including a noncubic metallic phase above 48~GPa and unusually large unit cells at lower pressures. This complexity, as well as the richness of strain-induced phases in BFO thin films \cite{Dieguez2011}, has motivated numerous studies of Bi-based perovskites.

Here, we are interested in understanding bismuth manganite, BiMnO$_3$ (BMO). Although under focus for some time as a potential multiferroic, BMO has been less studied experimentally, mainly because of its more demanding synthesis conditions. Its structural complexity combines the peculiar 6$s^2$ electrons and four interacting mechanisms: electric polarity, octahedra tilts, magnetism and cooperative Jahn-Teller distortion associated with the Mn$^{3+}$ cation. Ferroelectricity in BMO has been a controversial issue over the past few years. The ferroelectric structure originally proposed for bulk polycrystalline BMO was contradicted by subsequent studies (Refs. \cite{Belik2012,Goian2012} and references therein). On the other hand, ferroelectricity has been reported in thin films \cite{Son2008,DeLuca2013} while {\it ab initio} calculations have found no trend for ferroelectricity under epitaxial strain \cite{Hatt2009}. Here, our complementary use of Raman scattering and synchrotron X-ray diffraction (XRD) on both powder and single crystal shows that BMO exhibits above 37 GPa a polar monoclinic phase with an unprecedented giant elongation for a perovskite under high-pressure. Also remarkable in the pressure evolution of BMO are the reduction of the Jahn-Teller distortion, which occurs through a different process than the model system LaMnO$_3$, and an insulator-to-metal (IM) phase transition at 53~GPa. 

BiMnO$_3$ powder and single crystals were synthesized at high pressure--high temperature using the flux method in a belt-type apparatus as described before \cite{Toulemonde2009}. All diffraction and Raman scattering experiments were performed at room temperature in diamond-anvil cells, with helium as pressure transmitting medium. This is crucial to minimize nonhydrostatic stresses as they are known to potentially modify phase sequences in perovskites with multiple instabilities \cite{Guennou2011}. X-ray diffraction experiments were performed on the ID09A beam line at the ESRF ($\lambda=0.4144$~\AA, beam size $\approx$~20~$\upmu$m). Unpolarized Raman spectra were recorded on a Jobin-Yvon Labram spectrometer with a low-frequency cutoff at 100~cm$^{-1}$ and an exciting laser line at 633~nm. We have verified that all phase transitions are reversible.

\begin{figure}[ht]
\begin{center}
\includegraphics[width=0.48\textwidth]{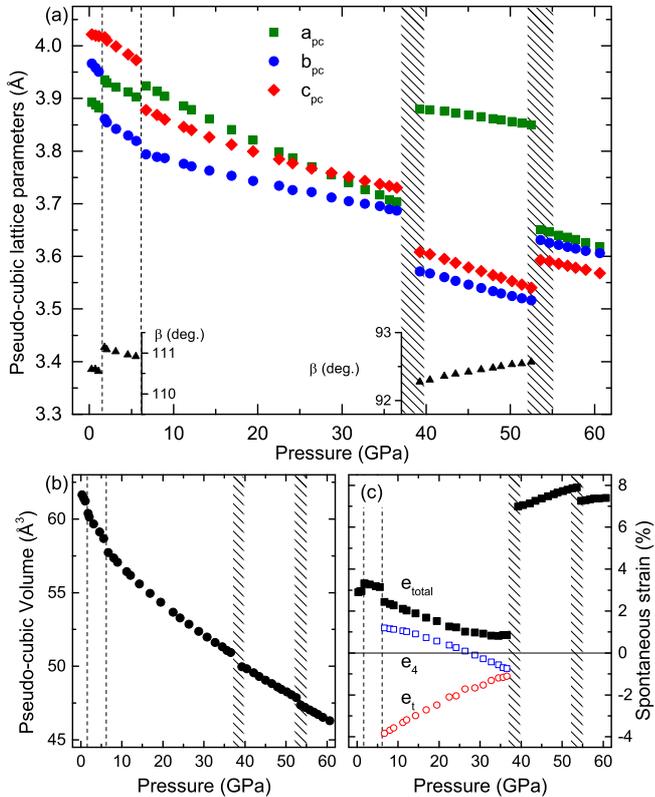}
\caption{(a) Lattice parameters, (b) volume and (c) spontaneous strains as a function of pressure (single crystal data). The hatched regions show phase coexistence. The powder data (not shown), follow the same evolution, albeit with much broader coexistence regions for all transitions. Since the structure above 53~GPa could not be determined with certainty, we estimated the volume by a Lebail fit of the integrated single crystal pattern with a triclinic cell (see values in the supplementary material).}
\label{fig:param}
\end{center}
\end{figure}

In the following we sequentially discuss the different parts of the pressure evolution which we divide into four parts: a low-pressure regime from ambient pressure to 6~GPa, the wide midpressure range from 6 to 37~GPa, the first high-pressure regime from 37 to 53~GPa, and a final regime above 53~GPa. The sequence of five phases is summarized in Fig. \ref{fig:param} where we show the evolution of the lattice parameters (a) and volume per formula unit $V/Z$ (b) with pressure. Discontinuities in the volume evolution show that all transitions are first order. Note that such a rich sequence of multiple phase transitions is unusual and only the similar Bi-perovskite BFO presents six phase transitions, while the vast majority of perovskites investigated so far presents none (CaTiO$_3$ \cite{Guennou2010a}), one (SrTiO$_3$ \cite{Guennou2010}, $R$MnO$_3$ \cite{Loa2001,Oliveira2012}) or two phase transitions (PbTiO$_3$ \cite{Janolin2008}). 

In the low-pressure regime, both Raman spectroscopy and XRD reveal two phase transitions at 1 and 6~GPa (see supplementary material). In agreement with literature data \cite{Belik2009a,Kozlenko2010}, we confirm the phase sequence $C2/c\rightarrow P2_1/c\rightarrow Pnma$. Both the $C2/c$ and $P2_1/c$ monoclinic structures have $Z=8$ and unit cell axes $a\approx a_c\sqrt{6}$, $b\approx a_c\sqrt{2}$, $c\approx a_c\sqrt{6}$ and $\beta\approx 110^\circ$ (with $a_c$ = equivalent cubic perovskite lattice parameter). The first transition is marked by the loss of the $C$-centering of the cell and the swapping of the $a$ and $b$ lattice parameters revealing a rearrangement in the $(a,b)$ plane. Unfortunately, twinning of the single crystal made a full structural refinement impossible. Interestingly, no phase coexistence is observed in the single crystal experiment. In contrast, the transitions in the powder are characterized by large regions of phase coexistence (first transition: between 1 and 4 GPa; second transition: between 5.7 and 7.7 GPa). This suggests that the wide phase coexistence observed here and in previous experiments \cite{Belik2009a,Kozlenko2010} is not of intrinsic origin but likely originates only from stress between grains, which in turn underlines the importance of single crystal experiments.

The midpressure regime is characterized by the orthorhombic $Pnma$ structure which remains stable between 6 and 37~GPa. This evolution is particularly interesting as an example of a Jahn-Teller (JT) distorted orthorhombic $Pnma$ perovskite over a large pressure range. Generally speaking, experimental and theoretical studies of JT systems \cite{Loa2001,Baldini2011,Ramos2011,Zhou2011a,Chen2012,Oliveira2012,He2012}, confirm the intuition that the JT is reduced under pressure, but the details of this evolution and its potential total suppression have been debated \cite{Baldini2011,Ramos2011,He2012}. In the model compound LaMnO$_3$ in particular, it was first claimed that the JT was completely suppressed above 18 GPa \cite{Loa2001} before further measurements by Raman and x-ray absorption spectroscopy suggested a persistence of the JT distortion up to the highest pressure preceding the IM transition at 34~GPa \cite{Baldini2011,Ramos2011}. 

Attempts at refining the crystal structure from the powder diffraction pattern at low pressure ($\approx$10~GPa) confirmed that BMO shows distortions typical of orbital ordering \cite{Belik2009a,Kozlenko2010} but did not enable us to extract reliable evolutions of the oxygen positions over a large enough pressure range. On the other hand, an analysis of spontaneous strains defined with respect to a cubic perovskite with lattice constant $a_0=(V/Z)^{1/3}$ can be done from the knowledge of the lattice parameters alone and gives information on the evolution of JT distortion. Following previous studies \cite{Carpenter2009,Carpenter2009a}, the relevant quantities are the shear strain $e_4$ and tetragonal strain $e_t$ (see supplementary material). Strains can in principle reflect the evolution of both tilt angles and cooperative JT. However, similarly to BFO, where strain in the $Pnma$ phase is due to tilts alone and shows virtually no pressure dependence \cite{Guennou2011a}, the tilt angles in BMO can be expected to vary little with pressure, so that the evolution of strains essentially reflects rearrangements of the orbital ordering. The absence of anomaly between 6 and 37~GPa excludes a $Pnma$ (tilt+JT)$\rightarrow Pnma$ (tilt only) transition. The general decrease of the total strain indicates that the distortion is strongly reduced, but we cannot ascertain whether it can be considered completely suppressed at 37~GPa. We note however that at this pressure we have $e_4<0$ (i.e. $c>a$), which is characteristic of perovskites with low tilt angles (resp. a tolerance factor close to 1), a shear distortion of BO$_6$ octahedra and a coordination mismatch at the $A$ site that is known to drive the system to a first-order transition upon substitution with larger $A$ cations, or an increase in temperature or hydrostatic pressure \cite{Arulraj2005,Shibasaki2005,Zhou2011}.

\begin{figure}[th]
\begin{center}
\includegraphics[width=0.48\textwidth]{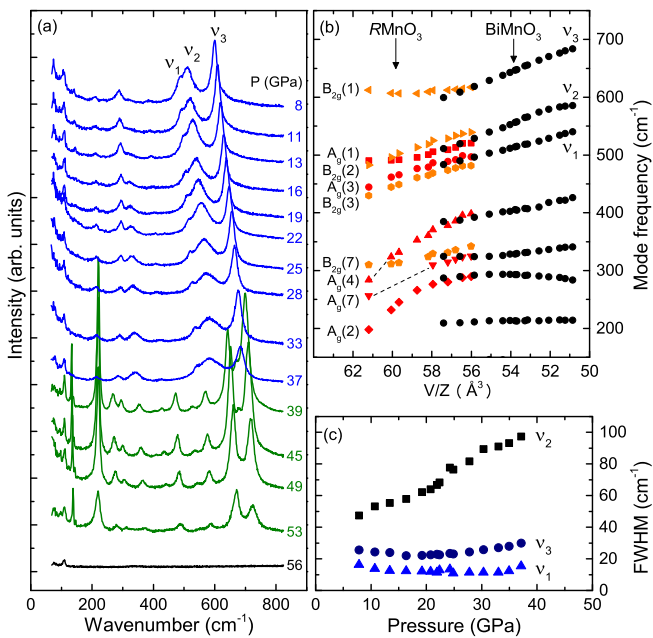}
\caption{(a) Selection of Raman spectra from 8 to 56~GPa. (b) Position of the main Raman bands as a function of volume per formula unit $V/Z$ for BiMnO$_3$ under pressure in the $Pnma$ phase and $R$MnO$_3$ at ambient conditions, taken from Ref. \cite{Iliev2006}. (c) Evolution of the bandwidths for the three Raman modes related to JT distorsion.}
\label{fig:Raman}
\end{center}
\end{figure}

The evolution of the Raman spectrum in the $Pnma$ phase is shown in Fig. \ref{fig:Raman}(a), and can be used to follow the evolution of the MnO$_6$ octahedra on a local scale, as it has been done on LaMnO$_3$ and related compounds \cite{Loa2001,Baldini2009,Baldini2011}. This requires an assignment of the Raman modes in BMO, which can be done by comparison with previous work on $Pnma$ rare-earth manganites $R$MnO$_3$ \cite{Iliev2006}. For this, we compare in Fig. \ref{fig:Raman}(b) the Raman modes in orthorhombic BMO at high pressure with those of $R$MnO$_3$ at ambient conditions by plotting the positions of the Raman bands as a function of $V/Z$. The remarkable overlap shows that the mode assignment can safely be transferred from $R$MnO$_3$ to BMO. We note in particular the nonlinear behavior of the A$_g$(7) and A$_g$(2) modes, which can be accounted for by mode coupling, like in the $R$MnO$_3$ series \cite{Iliev2006}. The tilt modes A$_g$(2) and A$_g$(4) can be identified as well, and their relatively weak volume dependence supports a limited evolution of the tilt angles under pressure. The modes in the 450--750~cm$^{-1}$ range labelled $\nu_{1,2,3}$ are associated to breathing of the MnO$_6$ octahedra and are relevant for a discussion of the JT distortion. $\nu_3$ can be assigned to B$_{2g}$(1); the assignment of $\nu_1$ and $\nu_2$ is more delicate since they might contain several overlapping bands (bending and antisymmetric stretching modes) that we cannot disentangle here. $\nu_1$ and $\nu_3$ show a linear volume dependence and their widths decrease slightly before increasing (Fig. \ref{fig:Raman}(c)). $\nu_2$ on the other hand shows a nonlinear behavior and a very pronounced broadening. This overall signature is significantly different from previous observations on LaMnO$_3$, where the main feature was reported to be the splitting of the high-frequency breathing B$_{2g}$(1) mode into two distinct peaks with a gradual intensity transfer from one to the other. This feature was then interpreted as the signature of a coexistence of distorted and undistorted (or less distorted) octahedra \cite{Baldini2009,Baldini2011}. Here, we conclude that the reduction of the JT distortion proceeds through a continuous displacivelike process in BMO and does not show the coexistence of different MnO$_6$ octahedra proposed for LaMnO$_3$. 

\begin{figure}[ht]
\begin{center}
\includegraphics[width=0.48\textwidth]{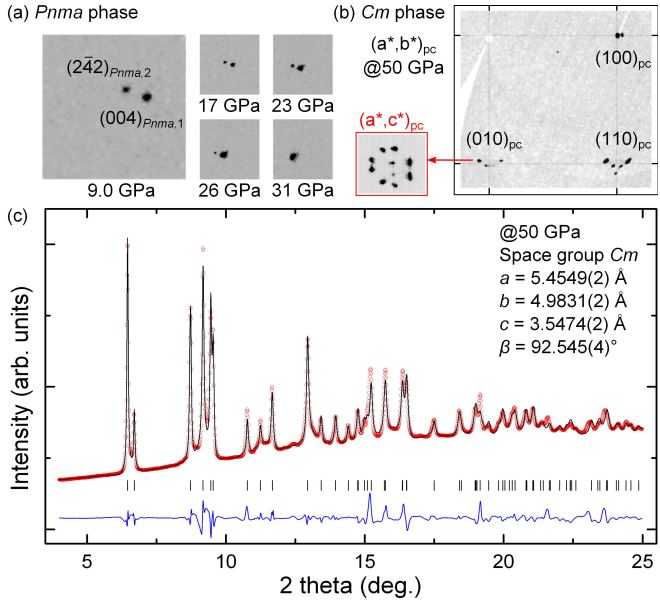}
\caption{(a) Illustration of the detwinning of the crystal under pressure in the $Pnma$ phase; the indices 1 and 2 refer to the two different domains. (b) Selected region of the reciprocal $(a^*,b^*)_{pc}$ plane (pseudo-cubic settings) showing the splittings associated with the domain structure in the $Cm$ phase. The zoom shows a map of the $(010)_{pc}$ reflection in the $(a^*,c^*)_{pc}$ plane ($(110)+(001)$ in the monoclinic settings). (c) Rietveld refinement (calculated profile in black line) of the experimental powder pattern (red open dots) at 50~GPa (see supplementary material for details).}
\label{fig:phase4}
\end{center}
\end{figure}

Between 37 and 39~GPa we observe a further phase transition revealed by the emergence of new peaks in the Raman spectrum (Fig. \ref{fig:Raman}(a)), and a very clear change in the diffraction pattern (Fig. \ref{fig:phase4}). Already qualitatively, we note in figure \ref{fig:phase4} a strikingly large splitting of the diffraction peaks, providing evidence for the sudden onset of a considerable distortion, which is highly unusual at such high pressure. For a better understanding of its origin we first consider the evolution of the domain structure in the single crystal pattern which is equally remarkable. Before the transition, the original twinning gradually disappears, similarly to BFO$_3$ \cite{Guennou2011a}, so that the crystal is in a single domain state between 30 and 37~GPa (Fig. \ref{fig:phase4}(a)). After the transition, the diffraction pattern shows an (at least) eight-variant ferroelastic domain structure (Fig. \ref{fig:phase4}(b)), which can only be expected for a monoclinic or triclinic structure. We also note the absence of superstructure peaks, which indicates the absence of strong octahedral tilts. A monoclinic unit cell with $Z=2$, $a\approx a_c\sqrt{2}$, $b\approx a_c\sqrt{2}$, $c\approx a_c$ and $\beta\approx 92.5^\circ$ was indeed identified to index the single crystal and powder diffraction patterns, with reflection conditions leaving only $C2$, $Cm$ and $C2/m$ as possible space groups. The quality of tentative Rietveld refinements on the powder patterns using this cell can be considered reasonable given the very high pressure (Fig. \ref{fig:phase4}(c) and Supplemental Material), but did not enable us to identify positively the most suitable space group. However, the inspection of possible atomic positions in a $C2/m$ cell shows that the atoms all have to occupy special positions in the unit cell and sit at an inversion center so that no Raman active mode would be expected at all in this structure. Therefore, this space group can be safely ruled out and the polar $C2$ and $Cm$ space groups remain as possible candidates. In both of these space groups, a total of 27 Raman-active modes can be expected, which is more than the 24 Raman active mode of the $Pnma$ phase, and therefore in qualitative agreement with the emergence of new Raman modes at the transition. Although we cannot discriminate between $Cm$ and $C2$ on the basis of our XRD and Raman data we note that i) $Cm$ phases with the same unit cell axes have been reported in related Bi-based perovskites \cite{Yusa2009,Belik2012a}, and ii) there is no relation to the $C2$ structure considered for BMO at ambient conditions because of different unit cell axes. Regardless of the difficulty in assigning the final space group, the crucial point is that both are polar space groups that allow ferroelectricity. This polar character can now explain the strongly elongated cell and the very large distortion, which we plot in Fig. \ref{fig:param}(c) by the total spontaneous strain $e_{\mathrm{total}}$ calculated with Aizu's definition for all phases \cite{Salje1993}. It shows a jump at the $Pnma\rightarrow Cm(C2)$ transition and reaches 8\% at 53~GPa, which is considerably larger than the distortions seen at lower pressure. It also exceeds the distortion reported for ferroelectric BMO thin film ($c/a=1.03$, $e_{\mathrm{total}}\approx 2$\%) \cite{Son2008}, and even for archetypal PbTiO$_3$ at ambient conditions ($c/a=1.064$, $e_{\mathrm{total}}\approx 5$\%), in line with the general view that a high $c/a$ ratio is a good fingerprint for ferroelectricity. 

The emergence of a polar phase at very high pressure is a remarkable if not unique feature for a perovskite. It contradicts the traditional expectation that high pressure suppresses ferroelectricity \cite{Samara1975} and sheds light on the conditions for such strongly elongated polar phases to occur in Bi-based perovskites. More recently, the reentrance of ferroelectricity at very high pressure has been theoretically predicted \cite{Kornev2005,Kornev2007,Bousquet2006}, but through a mechanism that produces only very small distortions driven by the $B$ cation, which we rule out here. Given the magnitude of the distortion, it appears natural to relate our observation to the classical ferroelectric activity of bismuth. This hypothesis gains support from the comparison with rare-earth manganites $R$MnO$_3$, which have the orthorhombic $Pnma$ structure at ambient conditions but do not exhibit any transition to a polar phase under pressure \cite{Loa2001,Oliveira2012}. Comparison with BFO is more delicate. BFO does exhibit multiple phase transitions under pressure, but no similar polar phase. Strongly distorted polar phases have been observed and predicted in BFO, but in conditions of epitaxial strain \cite{Bea2009} or negative hydrostatic pressure \cite{Ravindran2006} that are not readily comparable to hydrostatic compression. At 35~GPa, just before the transition, BFO and BMO have comparable volumes, but differ in at least two ways: the presence of a JT active $B$ cation and the amplitudes of tilt angles revealed by the different strain states. The mere presence of the JT active cation does not seem to be necessary considering that, in most Bi-based perovskites where a $Cm$ polar phase has been observed, the $B$ cation is not JT active (Fe$^{3+}$, Cr$^{3+}$, Al$^{3+}$, Ga$^{3+}$) \cite{Belik2012a}. However, the importance of the strong reduction, if not total suppression, of the cooperative JT distortion by hydrostatic pressure is supported by the observations that in BiMn$_{1-x}$Ga$_x$O$_3$, the $Cm$ polar phase is only stable for $x\ge 0.66$, a composition for which the long-range cooperative JT distortion can be expected to be completely suppressed by analogy with the LaMn$_{1-x}$Ga$_x$O$_3$ system \cite{Blasco2002}. We anticipate that the suppression of JT distortion can be obtained by a combination of pressure and chemical substitution, and that such a polar phase should be therefore stable in the $x$--$P$ phase diagram of BiMn$_{1-x}$Ga$_x$O$_3$ in a much broader region than previously expected \cite{Belik2012a}. This factor might as well play a role in the stabilization of a ferroelectric phase by epitaxial strain in thin films \cite{DeLuca2013}. However, a deep understanding of the interaction between JT distortion, octahedra tilts and ferroelectricity under pressure will require more theoretical and experimental investigations.

The last and again drastic phase transition in the pressure evolution of BiMnO$_3$ occurs between 52 and 55~GPa and is characterized by the vanishing of the Raman spectrum and a marked change in the diffraction pattern. The vanishing of the Raman signature on its own provides conclusive evidence for a phase transition, but it does not allow elucidating its nature and two scenarios can be envisaged. (i) BMO proceeds from the $Pnma$ to the ideal perovskite $Pm\overline 3m$ structure. In this scenario, the loss of the Raman spectra is explained by the fact that Raman scattering is forbidden by symmetry. (ii) BMO undergoes an insulator-to-metal (IM) transition, similarly to BFO, so that the loss of the Raman spectrum can be explained by the strongly reduced penetration depth of the laser in the metallic phase. Our XRD data allow us to rule out a $Pm\overline 3m$ cubic cell: the single crystal pattern (not shown) shows a threefold splitting of the $(001)$ reflection and a fourfold splitting of the $(011)$ reflection (at least), indicating that the highest possible symmetry is monoclinic, which again contrasts with the high-pressure phase of BFO. The exact structure determination, as well as the mechanism for the IM transition, will require further investigations. 

In summary, we have shown using a combination of high-pressure Raman spectroscopy, single crystal and powder diffraction that BiMnO$_3$ shows a remarkable structural sequence under high pressure $C2/c\rightarrow P2_1/c\rightarrow Pnma\rightarrow Cm (C2)\rightarrow X$. It exhibits an original evolution of the cooperative Jahn-Teller distortion in its $Pnma$ phase, a polar monoclinic phase at high pressure and an insulator-to-metal transition. BiMnO$_3$ therefore appears as a useful case study for a theoretical understanding by {\it ab initio} calculations of the special chemistry of Bi-based perovskites. Experimental investigations of other such Bi-based perovskites in the same pressure range are also desirable. 


\clearpage

\begin{widetext}
\begin{center}
{\Large Supplemental material}
\end{center}

\begin{figure}[ht]
\begin{center}
\includegraphics[width=0.50\textwidth]{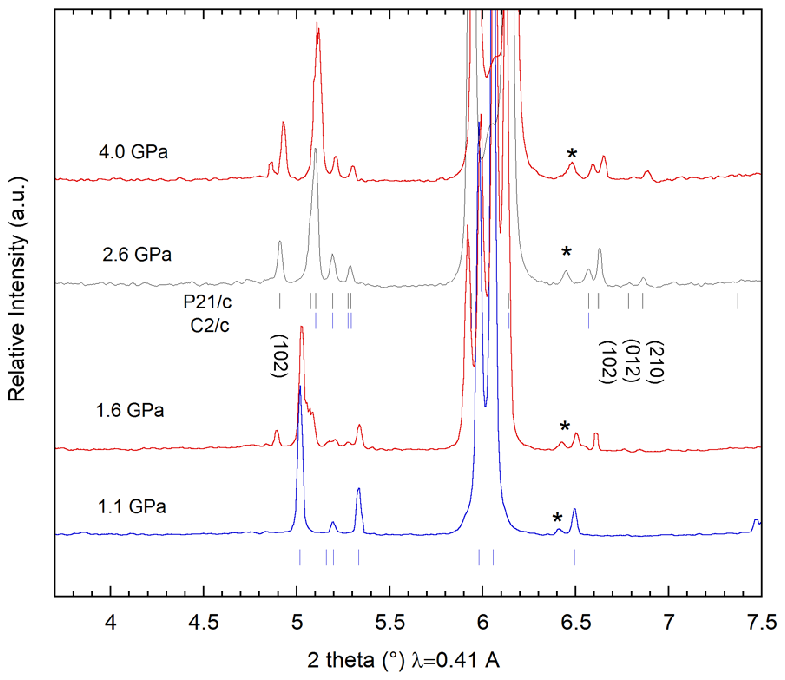}
\includegraphics[width=0.40\textwidth]{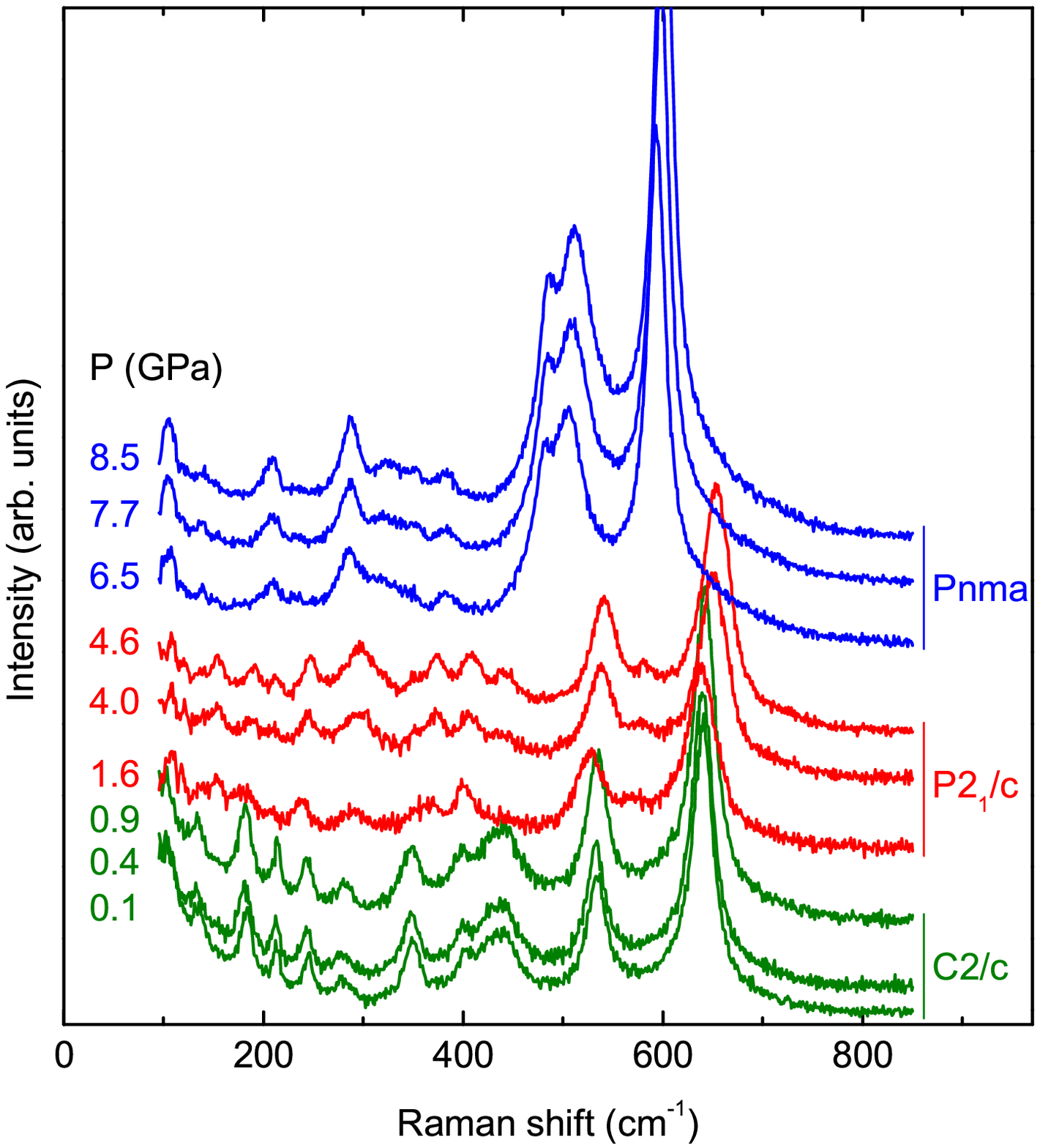}
\caption{Left: $C2/c\rightarrow P2_1/c$ transition observed in the powder diffraction pattern. The ticks indicate the pics predicted for the two phases and the asterisk marks a peak originating from a small fraction ($\approx$~2\%) of bismuthite impurity phase. Right: Raman spectra of the low-pressure phase transitions.}
\label{fig:XRD1}
\end{center}
\end{figure}

\clearpage

\begin{table}[ht]
\begin{center}
\caption{Lattice parameters from the single crystal data.}
\renewcommand{\arraystretch}{1}
\begin{tabular}{c c c c c c c c}
\hline\hline
$P$ [GPa]	& $a$	[\AA] & $b$	[\AA]	& $c$	[\AA]	& $\alpha$ [deg.] & $\beta$ [deg.] & $\gamma$ [deg.] & $V$\\
\hline
$C2/c$\\
0.29(5)		& 9.5348(6)		& 5.6092(2) &	9.8515(16)	&& 110.597(8)		&& 493.21(10)	\\
0.74(10)	& 9.5234(11)	& 5.5976(4)	& 9.8465(13)	&& 110.598(7)		&& 491.34(7)	\\
1.11(6)		& 9.5111(27)	& 5.5872(2)	& 9.8430(19)	&& 110.547(11)	&& 489.78(11)	\\
\\											
$P2_1/c$\\											
1.82(30) 	&	9.6394(7)		& 5.4608(1)	& 9.8381(10)	&&111.136(4)	&& 483.03(8)	\\
2.08(7)		& 9.6270(7)		& 5.4512(2)	& 9.8234(10)	&&111.081(5)	&& 481.01(8)	\\
3.16(11)	&	9.6060(7)		& 5.4339(2)	& 9.7958(10)	&&111.030(5)	&& 477.26(8)	\\
4.60(25)	&	9.5843(12)	& 5.4158(2)	& 9.7579(16)	&&110.955(7)	&& 472.99(12)\\
5.56(12)	&	9.5609(14)	& 5.4013(2)	& 9.7318(17)	&&110.910(8)	&& 469.47(14)\\
\\											
$Pnma$\\											
6.73(21)	&	5.5495(1)	&	7.5881(7)	 	&5.4840(1)	&&&& 	230.93(3)	\\
8.02(18)	&	5.5344(1)	&	7.5775(5)	 	&5.4713(2)	&&&&	229.45(3)	\\
8.95(10)	&	5.5206(1)	&	7.5651(9)	 	&5.4590(1)	&&&&	227.99(4)	\\
11.21(10)	&	5.4963(1)	&	7.5487(5)	 	&5.4386(2)	&&&&	225.65(3)	\\
12.11(12)	&	5.4856(1)	&	7.5423(6)	 	&5.4303(2)	&&&&	224.67(3)	\\
14.25(12)	&	5.4603(1)	&	7.5227(7)	 	&5.4114(2)	&&&&	222.28(3)	\\
16.89(13)	&	5.4315(1)	&	7.5046(8)	 	&5.3921(2)	&&&&	219.79(4)	\\
19.46(9)	&	5.4041(2)	&	7.4852(8)	 	&5.3733(2)	&&&&	217.35(4)	\\
22.51(21)	&	5.3719(3)	&	7.4687(40) 	&5.3520(4)	&&&&	214.73(10)\\
24.15(31)	& 5.3551(3)	&	7.4521(40)	&5.3413(4)	&&&&	213.15(10)\\
26.38(11)	& 5.3320(3)	&	7.4442(69)	&5.3273(5)	&&&&	211.45(18)\\
28.76(12)	& 5.3099(4)	&	7.4234(70)	&5.3153(5)	&&&&	209.51(18)\\
30.79(27)	& 5.2904(4)	&	7.4095(71)	&5.3048(5)	&&&&	207.95(18)\\
32.71(11)	& 5.2709(4)	&	7.4001(71)	&5.2936(5)	&&&&	206.48(18)\\
34.49(27)	& 5.2562(4)	&	7.3908(78)	&5.2859(5)	&&&&	205.34(20)\\
35.61(8)	& 5.2437(5)	&	7.3789(92)	&5.2799(6)	&&&&	204.29(23)\\
36.51(33)	& 5.2371(6)	&	7.3742(91)	&5.2756(5)	&&&&	203.74(24)\\
\\											
$Cm$\\											
39.25(10)	& 5.4869(3)	& 5.0509(2)	& 3.6085(3)	&& 92.274(11) &&	99.926(17)\\
40.46(10)	& 5.4852(3)	& 5.0454(2)	& 3.6038(3)	&& 92.305(12) &&	99.656(17)\\
42.15(36)	& 5.4809(3)	& 5.0350(2)	& 3.5954(3)	&& 92.358(12) &&	99.134(17)\\
43.50(9)	& 5.4758(3)	& 5.0251(2)	& 3.5874(3)	&& 92.390(13) &&	98.628(18)\\
45.02(7)	& 5.4713(3)	& 5.0154(2)	& 3.5792(4)	&& 92.421(15) &&	98.126(18)\\
46.55(7)	& 5.4665(3)	& 5.0063(2)	& 3.5717(4)	&& 92.454(16) &&	97.657(19)\\
47.94(3)	& 5.4615(2)	& 4.9974(2)	& 3.5646(4)	&& 92.479(15) &&	97.197(18)\\
48.90(9)	& 5.4582(2)	& 4.9916(2)	& 3.5594(4)	&& 92.502(16) &&	96.882(18)\\
50.20(23)	& 5.4539(2)	& 4.9850(2)	& 3.5531(5)	&& 92.530(18) &&	96.505(19)\\
51.35(10)	& 5.4492(2)	& 4.9785(2)	& 3.5465(5)	&& 92.542(20) &&	96.117(21)\\
52.50(7)	& 5.4441(2)	& 4.9731(2)	& 3.5403(7)	&& 92.568(23) &&	95.752(24)\\
\\
$P\overline 1$\\
54.64(8) 	& 3.6460(2)	&	3.6258(3)	&	3.5908(6)	&	84.99(1)	&	90.47(2)	&	92.94(1)	&	47.224(13)\\
55.73(9) 	& 3.6405(2)	&	3.6216(3)	&	3.5862(5)	&	84.98(1)	&	90.44(2)	&	92.96(1)	&	47.036(11)\\
56.80(12) & 3.6363(2)	&	3.6181(3)	&	3.5821(5)	&	84.95(1)	&	90.44(2)	&	92.99(1)	& 46.880(11)\\
57.77(7) 	& 3.6319(2)	&	3.6147(4)	&	3.5787(7)	&	84.95(1)	&	90.46(2)	&	93.04(1)	&	46.732(15)\\
58.95(4) 	& 3.6255(2)	&	3.6102(3)	&	3.5747(7)	&	84.97(1)	&	90.43(2)	&	93.09(1)	&	46.540(14)\\
60.64(17) & 3.6185(2)	&	3.6059(3)	&	3.5681(7)	&	85.01(1)	&	90.35(2)	&	93.18(1)	&	46.307(14)\\
\hline
\hline
\end{tabular}
\end{center}
\end{table}

\clearpage

\begin{table}[ht]
\begin{center}
\renewcommand{\arraystretch}{2}
\caption{Definition of the pseudo-cubic lattice parameters and components of the spontaneous strain. In all cases, we have $a_0=(V/Z)^{1/3}$, and $\beta^*$ and $\beta_0^*$ are the reciprocal lattice angles for the real and reference cell respectively. For the $Pnma$ phase, the reference axes are taken along the pseudo-cubic axes, i.e. they are rotated by 45$^\circ$ in the $(a,c)_{Pnma}$ plane. $e_t$ is the symmetry adapted tetragonal strain, defined with the unique axis along the $b$ axis of the $Pnma$ phase. The total strain is computed as $e_\mathrm{tot}=\sqrt{\sum e_{ij}^2}$ for all phases.}
\begin{tabular}{c c c}
\hline\hline
\multicolumn{3}{c}{Pseudo-cubic lattice constants}\\
$C2/c$ and $P2_1/c$ & $Cm$ & $Pnma$ \\
\hline
$\displaystyle a_\mathrm{pc} = \frac{a}{\sqrt{6}}$ & $\displaystyle a_\mathrm{pc} = \frac{a}{\sqrt{2}}$ & $\displaystyle a_\mathrm{pc} = \frac{a}{\sqrt{2}}$ \\
$\displaystyle b_\mathrm{pc} = \frac{b}{\sqrt{2}}$ & $\displaystyle b_\mathrm{pc} = \frac{b}{\sqrt{2}}$ & $\displaystyle b_\mathrm{pc} = \frac{b}{2}$ \\
$\displaystyle c_\mathrm{pc} = \frac{c}{\sqrt{6}}$ & $\displaystyle c_\mathrm{pc} = c$ & $\displaystyle c_\mathrm{pc} = \frac{c}{\sqrt{2}}$ \\
\hline
\multicolumn{3}{c}{Spontaneous strains}\\
\multicolumn{2}{c}{Monoclinic phases} & Orthorhombic phase\\
\multicolumn{2}{c}{$\displaystyle e_1 = e_{11} = \frac{a_\mathrm{pc}}{a_0}-1$}  & $\displaystyle e_1 = e_{11} = \frac{1}{2}\left[\frac{a_\mathrm{pc}+c_\mathrm{pc}}{a_0}-2\right]$ \\
\multicolumn{2}{c}{$\displaystyle e_2 = e_{22} = \frac{b_\mathrm{pc}}{a_0}-1$}  & $\displaystyle e_2 = e_{22} = \frac{b_\mathrm{pc}}{a_0}-1$ \\
\multicolumn{2}{c}{$\displaystyle e_3 = e_{33} = \frac{c_\mathrm{pc}\sin\beta^*}{a_0\sin\beta_0^*}-1$} & $\displaystyle e_3 = e_{33} = e_1 $\\
\multicolumn{2}{c}{$\displaystyle e_5 = 2e_{13} = \frac{a_\mathrm{pc}\cos\beta_0^*}{a_0\sin\beta_0^*}-\frac{c_\mathrm{pc}\cos\beta^*}{a_0\sin\beta_0^*}$}
& $\displaystyle e_4 = 2e_{23} = \frac{a_\mathrm{pc}-c_\mathrm{pc}}{a_0}$\\
&& $\displaystyle e_t = \frac{1}{\sqrt{3}}(2e_2-e_1-e_3)$\\
\hline
\hline
\end{tabular}
\end{center}
\end{table}

\begin{table}[ht]
\begin{center}
\caption{Details for the Rietveld refinement at 50~GPa with the space group $Cm$. The $B$ were not refined. Agreement factors for this fit are $R_\mathrm{p}=14.8$, $R_\mathrm{wp}=15.2$, $R_\mathrm{exp}=18.08$, $\chi^2=0.704$. For the corresponding centrosymmetric model with space group $C2/m$, all atoms are fixed at their special position and the agreement factors are $R_\mathrm{p}=16.6$, $R_\mathrm{wp}=16.8$, $R_\mathrm{exp}=17.87$, $\chi^2=0.879$.}
\renewcommand{\arraystretch}{1.5}
\begin{tabular}{c c c c c c}
\hline\hline
\multicolumn{2}{c}{Space group} & $a$ [\AA] & $b$ [\AA] & $c$ [\AA] & $\beta$ [deg.] \\
\multicolumn{2}{c}{$Cm$} 	& 5.4549(2)		& 4.9831(2)	& 3.5474(2) 	& 92.545(4) \\
\hline
Site 	& Wyckoff position 	& $x$ 				& $y$ 			& $z$ 				& B (\AA$^2$) \\
Bi 		& 2a 								& 0.0068(51) 	& 0.0 			& 0.0062(6) 	& 0.5 \\
Mn 		& 2a 								& 0.5					& 0.0				& 0.5					& 0.7 \\
O1 		& 2a 								& 0.385(11) 	& 0.0				& 0.049(15) 	& 1.0 \\
O2 		& 4b 								& 0.169(8)  	& 0.218(7) 	& 0.4914(15) 	& 1.0 \\
\hline
\hline
\end{tabular}
\end{center}
\end{table}

\clearpage

\end{widetext}

\end{document}